\documentclass[aps,reprint]{revtex4-1}                  
             
\usepackage{graphicx}
\usepackage{amssymb}
\usepackage{amsmath}
\usepackage{subfigure,wrapfig}

\begin{document}

\title{Vortex flow around a circular cylinder above a plane}

\author{M.~N.~Moura }
\author{G.~L.~Vasconcelos}\email[Corresponding author. Electronic mail: ]{giovani@df.ufpe.br}

\affiliation{Laborat\'orio de F\'{\i}sica Te\'orica e Computacional, 
Departamento de F\'{\i}sica, Universidade Federal de Pernambuco,
50670-901, Recife, Brazil.}

\date{\today}

\begin{abstract}
The study of vortex flows in the vicinity of multiple solid obstacles is of considerable theoretical interest and practical importance. In particular, the case of flows past a circular cylinder placed above a plane wall has attracted a lot of attention recently. In this case, a stationary vortex is formed in front of the cylinder, in contradistinction to the usual case without the plane where a vortex pair is observed behind the cylinder. In the present work, we apply modern complex analysis techniques  to obtain the complex potential for the problem of one point-vortex placed in a uniform stream past a circular cylinder above a plane. A typical  streamline pattern is also shown.
\end{abstract}


\maketitle

\section{Introduction}
\label{intro}

The formation of vortices in viscous flows past cylindrical structures  is a problem of considerable theoretical interest and practical relevance for many applications \cite{sf}. For example, in the case of  a flow past a circular cylinder  a pair of counter-rotating vortices forms behind the cylinder at small Reynolds numbers. This vortex pair then goes unstable at higher Reynolds numbers and evolves into a von K\'arm\'an vortex street.  This system was first studied analytically by F\"oppl in 1913 \cite{foeppl}. Modelling the vortex flow in terms of  point vortices in an otherwise inviscid and irrotational flow, F\"oppl was able to find stationary configurations for a pair of vortices behind the cylinder and analyze their stability. (Part of  F\"oppl's original stability analysis was however  in error, as it has been pointed out by several authors \cite{tangaubry,us}.) Because it is amenable to analytical treatment, the point-vortex model is an important tool to study the basic aspects of vortex dynamics in real fluids.  In addition, the study of vortex phenomena directly from the Navier-Stokes equation is very costly computationally, which makes the point-vortex model even more attractive.

\begin{figure}
\includegraphics[width=8cm]{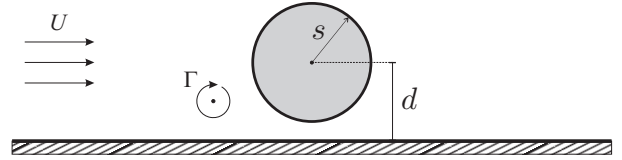}
\caption{Point vortex in a uniform stream past a circular cylinder above a plane.}
\label{fig:cylinder_wall}
\end{figure}

The problem of vortex flow in the presence of several solid boundaries (obstacles) is much more difficult, and theoretical studies for such cases are more sparse in the literature. One geometry of particular interest is that of a uniform flow past a circular cylinder placed above a planar solid wall, where vortices can form upstream of the cylinder \cite{lin}. Here  we wish to  investigate this problem in terms of a point-vortex model. More specifically, we consider  a point vortex of intensity $\Gamma$ placed in a uniform stream of velocity $U$ past a circular cylinder of radius $s$, whose center lies at a distance $d$ from a plane; see Fig.~\ref{fig:cylinder_wall}. The main objective of the present paper is to derive the complex potential for this system. A typical  streamline pattern of the flow will also be presented. 

Since the fluid domain for the case in study (Fig.~1) is not simply connected (nor can it be reduced to a simply connected region by symmetry arguments), we need to consider conformal mappings between multiply connected domains  in order to compute the complex potential for the flow.  More specifically, we shall make use of  a mathematical formalism recently developed by Crowdy and Marshall \cite{crowdymarshall}, which is based on the so-called Schottky-Klein prime function, as discussed next.

\section{Methodology}

\begin{figure}
\includegraphics[width=8cm]{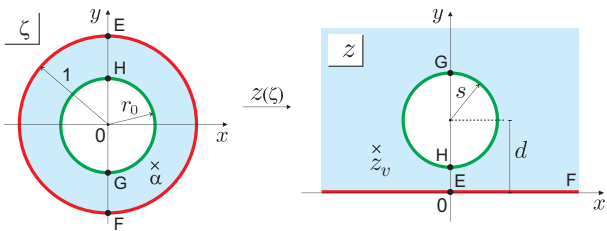}
\caption{Circular domain in the auxiliary complex $\zeta$-plane (left) and physical domain in the complex $z$-plane (right).}
\label{fig:mapscheme}
\end{figure}

We assume that the fluid is inviscid,  irrotational, and incompressible, so that the fluid velocity $\vec{v}(x,y)$ is given by the gradient of a potential function:  $\vec{v}=\vec{\nabla}\phi$, where  the velocity potential $\phi(x,y)$ obeys Laplace equation:   $\nabla^2\phi=0$. Our main goal here is to compute the complex potential $w(z)=\phi(x,y)+i\psi(x,y)$, where  $\psi$ is the so-called stream function, for the problem illustrated in Fig.~1.  To this end, let us introduce the conformal mapping $z(\zeta)$ from an annular region, $r_0<|\zeta|<1$, in the auxiliary complex $\zeta$-plane onto the fluid domain  in the complex $z$-plane, where  the unit circle $|\zeta|=1$ is mapped to the plane boundary and the inner circle  $|\zeta|=r_0$ is mapped to the cylinder. Furthermore, the points $\zeta=i$ and $\zeta=-i$ are mapped to $z=0$ and $z=\infty$, respectively; see Fig.~\ref{fig:mapscheme}.

It is not difficult to see that the function that enacts the desired mapping is given by
\begin{equation}
z(\zeta)=-i\sqrt{d^2-s^2}\,\left(\frac{\zeta-i}{\zeta+i}\right),
\label{cmap1v}
\end{equation}
whose inverse is
\begin{equation}
\zeta(z)=-i \frac{z-i\sqrt{d^2-s^2}}{z+i\sqrt{d^2-s^2}} .
\label{incmap1v}
\end{equation}
The radius  $r_0$ of the inner circle in the $\zeta$-plane is related to the physical parameters $s$ and $d$ by the following expression: 
\begin{equation}
r_0=\frac{1-\sqrt{\frac{d-s}{d+s}}}{1+\sqrt{\frac{d-s}{d+s}}} ,
\label{r0}
\end{equation}
as can be easily verified.

Before presenting the complex potential for the problem above, it is instructive to recall \cite{saffman} that the complex potential for a point vortex of intensity $\Gamma$ located at position $z=z_v$ (in an unbounded domain) is given by
\begin{equation}
w(z)=\frac{\Gamma}{2 \pi i}\log\left(z-z_v\right) \:.
\label{wv}
\end{equation}
Similarly, for a vortex in the presence of a cylinder of radius~$a$, the complex potential can be easily obtained by virtue of  the Milne-Thomson circle theorem \cite{milne-thomson}, which yields
\begin{equation}
w(z)=\frac{\Gamma }{2\pi  i}\log \left[
\frac{z-z_v}{z- {a^2}/{\bar{z}_v}}\right].
\label{eq:ct}
\end{equation}
Here the term in the numerator comes from the contribution to the complex potential from the vortex itself, see Eq.~(\ref{wv}); whereas the term in the denominator corresponds to the  contribution from the vortex image (inside the cylinder), which is necessary to enforce the boundary condition that the cylinder surface be a streamline of the flow.

For the geometry shown in Fig.~1, where in addition to the cylinder there is an extra boundary (the plane $y=0$), the circle theorem is no longer of help, and one has to resort to an alternative approach in order to compute the contribution from the infinite set of vortex images (both inside the cylinder and below the plane). In this case, it is more convenient to compute the complex potential  in the auxiliary $\zeta$-plane and then transform it back to the $z$-plane.  Indeed,  it can be shown \cite{crowdymarshall,crowdynotas} that for a point vortex of unit intensity, located at position $\zeta=\alpha$ in a circular domain in the $\zeta$-plane, the complex potential $w_v(\zeta,\alpha)$ is given by
\begin{equation}
w_v(\zeta,\alpha)=\frac{1}{2 \pi i} \log \left[\frac{\omega(\zeta,\alpha)}{\left|\alpha\right|\omega\left(\zeta,\frac{1}{\bar{\alpha}}\right)}\right] \:,
\label{G0}
\end{equation}
where $\omega(\zeta,\alpha)$ is the so-called Schottky-Klein prime function, which encodes the geometry of the circular domain. 

For the particular geometry shown in Fig.~2, the   Schottky-Klein prime function $\omega(\zeta,\alpha)$ can be expressed in a simple form:
\begin{equation}
\omega(\zeta,\alpha)=-\frac{\alpha}{C} P\left(\frac{\zeta}{\alpha},r_0\right),
\label{omega}
\end{equation}
where
\begin{equation}
C=\prod_{n=1}^{\infty}\left(1-r_0^{2n}\right)
\label{constC}
\end{equation}
and 
\begin{equation}
P(x,y)=(1-x)\prod_{n=1}^{\infty}\left(1-y^{2n}x\right)\left(1-y^{2n}x^{-1}\right).
\label{pfunc}
\end{equation}
From the definition of the complex potential $w_v(\zeta,\alpha)$ given in Eqs.~(\ref{G0})-(\ref{pfunc}), one can verify that it  satisfies the appropriate boundary conditions in the $\zeta$-plane, namely, that the unit circle and the inner circle are both streamlines of the flow.

The function $P(x,y)$ given in Eq.~(\ref{pfunc})  is related to  the first Jacobi theta function  $\vartheta_1$, which appears in the theory of  elliptic functions \cite{crowdyjacobi}.  Indeed, it is possible to derive  an expression for the complex potential $w_v(\zeta,\alpha)$ entirely  within the framework of elliptic functions. The advantage of the method based on the Schottky-Klein prime function   is that it can  be rather easily extended  to two-dimensional vortex flows around an arbitrary number of obstacles. (This general problem is  however beyond the scope of the present work.)

Using the complex potential for a vortex of unit intensity given in Eq.~(\ref{G0}), it is possible to obtain the complex potential, $w_U(z)$, due to a uniform stream of velocity $U$ past a cylinder above a plane (with no other flow elements). In this case, it can be shown \cite{tese} that the corresponding complex potential, $w_U(\zeta)$, in the $\zeta$-plane is given by
\begin{equation}
w_U(\zeta)=-2 \pi U i\sqrt{d^2-s^2} \left.\left(\frac{\partial w_v}{\partial \bar{\alpha}}-\frac{\partial w_v}{\partial \alpha}\right)\right|_{\alpha=-i} \:,
\label{wuf2}
\end{equation}
with $w_v$ as in Eq.~(\ref{G0}). Notice that the derivatives in Eq.~(\ref{wuf2}) are evaluated at the point $\alpha=-i$, which is the point that is mapped to infinity by the conformal map $z(\zeta)$, i.e., $z(-i)=\infty$. In the next section we shall use Eqs.~(\ref{wuf2}) and (\ref{G0}) to construct the complex potential for the vortex flow illustrated in Fig.~1.

\section{Results and Discussion}

\begin{figure}
\includegraphics[width=8cm]{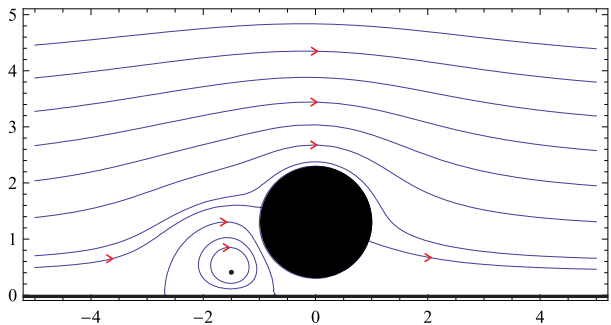}
\caption{Streamline pattern for the flow studied. The parameters chosen are  $\Gamma=-10$,   $U=1$, and $z_v=-1.5+0.5i$.}
\label{fig:cplotuf1vfull}
\end{figure}

The complex potential $w(z)$ for the problem of a vortex of intensity $\Gamma$ placed at position $z_v$ in a uniform stream in the presence of a cylinder and a plane boundary can be obtained by combining Eq.~(\ref{wuf2}) with  Eq.~(\ref{G0}), multiplied by $\Gamma$. This yields
\begin{equation}
w(z)=w_U(\zeta(z))+\Gamma w_v(\zeta(z),\zeta(z_v)) \:,
\label{fullpotential}
\end{equation}
where $\zeta(z)$ is the inverse mapping given in Eq.~(\ref{incmap1v}).
As is well known \cite{saffman}, the velocity field $\vec{v}=(u,v)$  can be found by simply taking the derivative of the complex potential:
\begin{equation}
u-iv=\frac{d w}{d z} \:.
\label{cvel}
\end{equation}
This yields a pair of differential equations for $\dot{x}=u(x,y)$ and $\dot{y}=v(x,y)$, which can be numerically integrated to generate the flow streamlines.  It is however more convenient to take an alternative approach, namely, perform a  contour plot of the stream function  $\psi(x,y)$, which can be easily obtained by taking the imaginary part of the complex potential (\ref{fullpotential}). Each level set  $\psi(x,y)=c$, with $c$ being a constant,  then yields a streamline of the flow. Figure \ref{fig:cplotuf1vfull} shows the streamline pattern obtained from the contour plot just described for the case when $U=1$, $\Gamma=-10$,  and the vortex is located at $z_v=-1.5 + 0.5 i$.

The streamline pattern shown in Fig.~\ref{fig:cplotuf1vfull} displays some interesting features.   Of particular interest is the formation of a recirculation zone around the vortex in front of the cylinder. A pattern similar to this was seen in the experiments on flows past a circular cylinder above a plane reported in Ref.~\cite{lin}. Notice that the recirculation zone acts as an ``obstacle'' to the oncoming flow,  deflecting part of the fluid over the top of the cylinder and thus reducing the flow though the gap. Note also that far above from the cylinder  the streamlines tend to be straight lines, meaning that far away from the cylinder one recovers the uniform flow imposed by the uniform stream, as expected. 

\section{Conclusions}

We computed the complex potential $w(z)$ for a point vortex in a uniform stream past a circular cylinder placed above a plane wall. From the imaginary part of $w(z)$, we  obtained the stream function of the flow whose level sets yield the streamlines.  In particular, we presented the  streamline pattern for the case when the vortex is placed upstream of the cylinder---a situation that is of interest to the experiments on viscous flows past a cylinder above a plane recently performed by Lin {\it et al.} \cite{lin}. The next step towards a more complete theoretical understanding of the system is to study the vortex dynamics in this geometry,  by allowing the vortex to move with the flow velocity    (excluding the vortex own contribution) at the vortex position. This analysis is currently in progress.

\begin{acknowledgments}
This work was supported in part by the Brazilian agencies CNPq and FACEPE.
\end{acknowledgments}

\end{document}